\documentclass[twocolumn,aps,showpacs,floatfix,prc]{revtex4-1}
\usepackage{graphicx}
\begin{document}

\title { Four pedagogic exercises }

\author{D.N. Basu}

\affiliation{Variable  Energy  Cyclotron  Centre, 1/AF Bidhan Nagar, Kolkata 700 064, India}
\email{dnb@veccal.ernet.in} 

\date{\today }

\begin{abstract}

    This article describes four innovative pedagogical exercises: (i) The expression for relation between cross sections in the laboratory and the centre of mass systems provided in text books assumes zero or low $Q$ values which needs to be corrected for the most general case when $Q$ value of the reaction is not negligible compared to the masses of interacting nuclei. The general expression derived here can be used for elastic and inelastic cases involving zero, low or even very high $Q$ values. (ii) The equation of oscillatory motion of a massive surface put horizontally on two wheels rotating with equal and opposite angular velocities is established. The time period of oscillation is related to the coefficient of dynamic friction between the surface and the wheels which facilitates the measurement of the coefficient of dynamic friction. (iii) The equation of motion of a fixed torque mass shedding vehicle moving against friction and its velocity at any instant of time are derived. This example is equivalent to motion of a mass shedding rocket moving by applying fixed force against atmospheric friction. (iv) The equation of the path of a missile directed at every instant of time towards a rectilinearly moving target and time taken to hit the target are derived. It provides equation of path for asteroid and comet destructing missiles as conceptualized by NASA projects.

\vspace{0.2cm} 


\end{abstract}

\maketitle

\noindent
\section{ Relation between laboratory and centre of mass cross sections for high $Q$ values }
\label{section1}

    In this short note the non-relativistic expression for the relation between cross sections in the laboratory and the centre of mass systems derived in text books \cite{Sc55} is re-derived. Let $m_1$ be the mass of the projectile incident with velocity $v_0$ and kinetic energy $E=\frac{1}{2}m_1v_0^2$ in the laboratory frame on a target of mass $m_2$ at rest. Let $m_3$ and $m_4$ be the masses of the interacting nuclei, respectively, after collision.

    The centre of mass system of interacting particles is defined as a frame of reference where the sum of the momenta of all interacting particles is zero. Therefore, in the centre of mass system incident projectile and the target nuclei moves with momenta which are equal in magnitude and opposite in direction and hence

\begin{equation}
 m_1(v_0-v_{cm}) = m_2 v_{cm} ~~ => ~~ v_{cm}=\frac{m_1v_0}{m_1+m_2}
\label{seqn1}
\end{equation}
\noindent
where $v_{cm}$ is the velocity of the centre of mass of the projectile-target system in the laboratory frame of reference. Obviously, $v_1=v_0-v_{cm}$ and $v_2=v_{cm}$ are the velocities of the projectile and the target, respectively, as observed in the centre of mass frame. Thus, the kinetic energy $E_{cm}$ in the centre of mass frame of reference is

\begin{equation}
 E_{cm} = \frac{1}{2}m_1 v_1^2 + \frac{1}{2}m_2 v_2^2 = \frac{m_2 E}{m_1+m_2}=\frac{1}{2} \mu v_0^2
\label{seqn2}
\end{equation}
\noindent
obtained using Eq.(1) where $\mu =\frac{m_1m_2}{m_1+m_2}$ is called the reduced mass of the projectile-target system and the magnitude of momenta of each particle $m_1(v_0-v_{cm})=m_2v_{cm}=\mu v_0$ are equal in this frame of reference. The energy released (or absorbed) $Q$ in the process of interaction is given by 

\begin{equation}
 (m_1+m_2)c^2 = (m_3+m_4)c^2 + Q
\label{seqn3}
\end{equation}
\noindent
which is always realised in the centre of mass frame of reference and $c$ is the velocity of light in vacuum. Let $v_3$ and $v_4$ be the velocities of scattered projectile like and recoiling target like nuclei in the centre of mass system and $\theta_0$ be the angle of scattering of the scattered projectile like nucleus of mass $m_3$ as observed in the laboratory system while $\theta$ be the corresponding angle as observed in the centre of mass system. Obviously, the two angles $\theta_0$ and $\theta$ are related by  

\begin{equation}
 tan\theta_0 = \frac{v_3sin\theta}{v_3cos\theta+v_2} = \frac{sin\theta}{cos\theta+\gamma}, ~~ \gamma=\frac{v_2}{v_3}
\label{seqn4}
\end{equation}
\noindent
obtained by constructing the velocity triangle. The following energy conservation equations in the centre of mass system 

\begin{eqnarray}
 \frac{1}{2}m_1 v_1^2 + \frac{1}{2}m_2 v_2^2 = E_{cm} ~~~~ {\rm before ~ collision} \nonumber \\
\frac{1}{2}m_3 v_3^2 + \frac{1}{2}m_4 v_4^2 = E_{cm} + Q ~~~~ {\rm after ~ collision}
\label{seqn5}
\end{eqnarray}
\noindent
can be used for evaluation of $\gamma$. In the centre of mass frame momenta of the two colliding nuclei are always equal in magnitude and opposite in direction at every instant of time and hence infinite time before and infinite time after collision as well, so that

\begin{equation}
  m_1v_1 =  m_2v_2, ~~  m_3v_3 =  m_4v_4
\label{seqn6}
\end{equation}
\noindent
Using Eq.(6) to eliminate $v_1$ and $v_4$ from Eq.(5), one obtains 

\begin{equation}
  \frac{E_{cm}}{E_{cm}+Q} = \frac{(m_1+m_2)}{(m_3+m_4)}\frac{m_2m_4}{m_1m_3}\Big(\frac{v_2}{v_3}\Big)^2
\label{seqn7}
\end{equation}
\noindent
and therefore

\begin{equation}
  \gamma = \frac{v_2}{v_3} = \Big[ \frac{m_1m_3}{m_2m_4}  \frac{(m_3+m_4)}{(m_1+m_2)} \frac{E_{cm}}{E_{cm}+Q} \Big]^{1/2}.
\label{seqn8}
\end{equation}
\noindent

    The relation between the differential cross sections measured in the laboratory system   $(\frac{d\sigma}{d\Omega})_{lab}$ and the centre of mass system $(\frac{d\sigma}{d\Omega})_{cm}$ is given by

\begin{equation}
  2\pi sin\theta_0 d\theta_0 (\frac{d\sigma}{d\Omega})_{lab} = 2\pi sin\theta d\theta (\frac{d\sigma}{d\Omega})_{cm}
\label{seqn9}
\end{equation}
\noindent
from the conservation of flux. Thus
    
\begin{equation}
   (\frac{d\sigma}{d\Omega})_{lab} = \frac{sin\theta d\theta }{sin\theta_0 d\theta_0} (\frac{d\sigma}{d\Omega})_{cm}
\label{seqn10}
\end{equation}
\noindent
Using Eq.(4), the above equation becomes 

\begin{equation}
  (\frac{d\sigma}{d\Omega})_{lab} = \frac{[1+\gamma^2+2\gamma cos\theta]^{3/2}}{|1+\gamma cos\theta|} (\frac{d\sigma}{d\Omega})_{cm}.
\label{seqn11}
\end{equation}
\noindent
The angle integrated cross sections are just the areas offered perpendicularly to the incident beam direction and therefore remain same in the laboratory and the centre of mass systems which means $\sigma_{lab}=\sigma_{cm}$.  

    The Eq.(8) provided above for $\gamma$ is exact and there is no approximation involved and is valid for the most general case when $Q$ value of the reaction is not negligible compared to the masses of the interacting nuclei. The term $\frac{(m_3+m_4)}{(m_1+m_2)}$ is missing in the corresponding expression of Ref.\cite{Sc55} which is valid only for $Q$ values much smaller than the masses involved and is therefore an approximate expression. For elastic scattering $Q=0$, $m_1= m_3$, $m_2=m_4$ and then $\gamma = \frac{m_1}{m_2}$ \cite{Go50} from the Eq.(8) above. 

\noindent
\section{ A laboratory method of measuring the coefficient of dynamic friction }
\label{section2}

    Measurement of the coefficient of dynamic (or kinetic) friction between wheel and road (or any surface) is often needed. A useful laboratory method for measuring the coefficient of dynamic (not rolling) friction is described here. First an equation is derived for the motion to show that the restoring force is proportional to the displacement of the plate (surface) from its equilibrium position. Then the time period of the simple harmonic motion (SHM) executed by the surface is related to the coefficient of dynamic friction between the surface and the wheels establishing the feasibility of this method for measuring the coefficient of dynamic friction.

    Let the centres of the axles of two wheels are $d$ distance apart and they are coupled by belt in such a manner that when motor of one wheel drives it at a constant speed (which means constant angular velocity), the belt forces the other wheel to move in the opposite direction and with identical speed as the former. The wheels are kept vertical, on which the plate of mass $M$ is placed whose dynamic friction coefficient against the given wheels is to be determined. Now at equilibrium condition, the plate rests symmetrically on the two wheels with its centre of mass at $\frac{d}{2}$ distance from each of the axles. If the plate is now displaced it will execute periodic oscillations. Let $x$ be the displacement at any instant of time $t$ and $R_1$ and $R_2$ be the normal reactions on the two wheels at that instant of time. Obviously, at equilibrium condition these two forces of normal reactions are equal. When the centre of mass of the plate is displaced by $x$, its distances from the two axles are $\frac{d}{2} + x$  and $\frac{d}{2} - x$, respectively. Balancing moments of the weight of the plate (acting vertically downward from its centre of mass) against reaction force (acting vertically upward and passing through axle) about the first and the second axles yield    

\begin{eqnarray}
 R_2 d =  (\frac{d}{2} + x) Mg, \nonumber\\
 R_1 d =  (\frac{d}{2} - x) Mg, 
\label{seqn1}
\end{eqnarray}
\noindent   
respectively, where $g$ is acceleration due to gravity. The frictional forces which are proportional to the normal reactions are, therefore, $\mu R_1$ and $\mu R_2$, respectively, and they are acting in the opposite directions horizontally but unequal in magnitude except for the mean position with $\mu$ being the coefficient of dynamic friction. Hence, the restoring force is  

\begin{equation}
 \mu R_1 - \mu R_2 = -\frac{2 \mu g M}{d} x
\label{seqn2}
\end{equation}
\noindent
where we have used Eqs.(1) for evaluating $R_1$ and $R_2$. The equation of motion (Newtonian force equation) is 

\begin{equation}
 M \frac{d^2x}{dt^2} = -\frac{2 \mu g M}{d} x
\label{seqn3}
\end{equation}
\noindent
which is the equation of a SHM. Its time period $T$ of oscillation is, therefore, given by

\begin{equation}
 T = 2 \pi \sqrt{\frac{d}{2 \mu g}}~. 
\label{seqn4}
\end{equation}
\noindent

    The coefficient of dynamic friction

\begin{equation}
 \mu = \frac{2 \pi^2 d}{g T^2}
\label{seqn5}
\end{equation}
\noindent
can now be measured by measuring the time period of oscillation $T$ and distance $d$ between the axles. It is obvious from the above equation that if the distance $d$ between the axles is increased then the time period of oscillation $T$ also increases. Obviously, for better accuracy, large separation between the axles is desired since it would not only lessen the errors involved in the measurement of separation distance $d$ but also reduces the errors in the time period measurements. 

\noindent
\section{ The equation of motion of a mass shedding vehicle }
\label{section3}

    A fixed torque vehicle applies fixed force since radii of its wheels are fixed. The problem of the motion of a mass shedding fixed torque vehicle moving against friction is equivalent to motion of a mass shedding rocket moving by applying fixed force against atmospheric friction provided frictional force is proportional to its variable mass which in turn is proportional to its length (atmospheric friction is likely to be proportional to its length since viscous drag depends upon the dimensions of the object). 

    Let $M$ be the mass of a vehicle (including the fuel and the load) at the start of the journey which is shedding mass at a constant rate of $r$ per unit time. The force that it applies to overcome frictional force and maintain its initial speed is $\mu gM$ which remains constant throughout as per the statement of the problem. The frictional force is assumed to be proportional to the weight of the vehicle at any instant with $\mu, g$ being the coefficient of friction and acceleration due to gravity, respectively. The net force at time $t$ is obviously $\mu gM - \mu g(M - rt) = \mu grt $ so that acceleration $\frac{dv}{dt}$ at time $t$ is given by  

\begin{equation}
 \frac{dv}{dt} =  \frac{\mu grt}{M-rt} 
\label{seqn1}
\end{equation}
\noindent
and hence velocity $v$ at time $t$ can be obtained from the following equation

\begin{equation}
 v = \int \frac{\mu grt}{M-rt} dt + c
\label{seqn2}
\end{equation}
\noindent
where $c$ is a constant of integration. Changing variable from $t$ to $x=rt$, one obtains

\begin{equation}
 v =  \frac{\mu g}{r}\int \frac{x}{M-x} dx + c. 
\label{seqn3}
\end{equation}
\noindent
The first term of the right hand side of the above expression can be integrated to provide 

\begin{equation}
 v =  \frac{\mu g}{r}[Mln(\frac{1}{M-x})-x] + c 
\label{seqn4}
\end{equation}
\noindent
which with the initial condition that when $t=0$, $v=u$ yields $c=u+\frac{\mu g}{r}MlnM$ where $u$ is the initial velocity of the vehicle. Therefore, the velocity $v$ at time $t$ of a fixed torque mass shedding vehicle moving against friction is 

\begin{equation}
 v = u + \frac{\mu g}{r}[Mln(\frac{M}{M-x})-x]
\label{seqn5}
\end{equation}
\noindent
where $x=rt$ is the amount of mass shed upto time $t$.

    In case of a rocket the term $\mu g$ can be replaced by some other constant $k$ and then it would mean that the engine of the rocket applies a constant force of magnitude $kM$ necessary to overcome atmospheric friction required initially and the atmospheric friction at any instant of time is proportional to its mass (which is proportional to its length) with $k$ as the constant of proportionality.

\noindent
\section{ The equation of path of an object tracking missile }
\label{section4}

    A missile is fired at a rectilinearly moving target when it is closest to it and hence the line joining the missile and the target is perpendicular to the straight line along which the target is moving with constant velocity $u$. The missile is moving with a constant speed of $v$ and is directed at every instant of time towards the moving target. Let $l$ be the distance of closest approach when the missile starts moving towards the target. After time $t$, let $s(x,y)$ be the length of the curvilinear path travelled by the missile which is moving with a constant speed $v$.  

    Since the missile is directed at every instant towards the rectilinearly moving target, the tangent on curvilinear path of the missile meets the target at any instant of time $t$ and therefore 

\vspace{-0.5cm}
\begin{eqnarray}
 vt = s(x,y) \nonumber \\
 ut = y + (l-x)\frac{dy}{dx}
\label{seqn1}
\end{eqnarray}
\noindent
assuming the line joining the missile and the target at $t=0$ as the x-axis which is perpendicular to the straight line y-axis along which the target is moving with constant velocity $u$ and position of the missile at $t=0$ to be the origin of the co-ordinate system. From the above two equations one obtains

\vspace{-0.5cm}
\begin{equation}
 y + (l-x)\frac{dy}{dx} =\frac{u}{v}s
\label{seqn2}
\end{equation}
\noindent
which can be differentiated with respect to $x$ to yield

\vspace{-0.5cm}
\begin{equation}
 \beta \frac{ds}{dx} = \frac{dy}{dx} + (l-x)\frac{d^2y}{dx^2} - \frac{dy}{dx} = (l-x)\frac{d^2y}{dx^2}
\label{seqn3}
\end{equation}
\noindent
where $\beta=\frac{u}{v}$ is a constant of motion. Putting $p=\frac{dy}{dx}$, the above equation reduces to

\vspace{-0.5cm}
\begin{equation}
 \beta \sqrt{1+p^2}=  (l-x)\frac{dp}{dx}
\label{seqn4}
\end{equation}
\noindent
since $ds^2=dx^2+dy^2$, $\frac{ds}{dx}=\frac{\sqrt{dx^2+dy^2}}{dx}=\sqrt{1+(\frac{dy}{dx})^2}=\sqrt{1+p^2}$. Therefore, solution of the above equation is \\

\vspace{-0.5cm}
\begin{equation}
  \int \frac{dp}{\sqrt{1+p^2}} = \beta \int \frac{dx}{(l-x)} + c_1
\label{seqn5}
\end{equation}
\noindent
where $c_1$ is a constant of integration. \\

    Evaluating the integrals on both sides provide 

\begin{equation}
  sinh^{-1}p = -\beta ln(l-x) + c_1.
\label{seqn6}
\end{equation}
\noindent
But at $t=0$, $x=0$ and $p=\frac{dy}{dx}=0$ since the x-axis is chosen in the direction perpendicular to the straight line y-axis along which the target is moving and position of the missile is chosen to be the origin of the co-ordinate system at $t=0$. This initial condition yields $c_1=\beta lnl$ and $sinh^{-1}p = \beta ln(\frac{l}{l-x})$. Therefore

\begin{equation}
  p = \frac{dy}{dx} =  sinh[\beta ln(\frac{l}{l-x})] = \frac{1}{2}[(\frac{l}{l-x})^\beta - (\frac{l}{l-x})^{-\beta} ]
\label{seqn7}
\end{equation}
\noindent
and integrating both the sides of the above equation yields

\begin{equation}
  y = \frac{1}{2}[\frac{l^{-\beta}(l-x)^{1+\beta}}{1+\beta} - \frac{l^{\beta}(l-x)^{1-\beta}}{1-\beta}] +c_2
\label{seqn8}
\end{equation}
\noindent
where $c_2$ is another constant of integration. Since at $t=0$, both $x=0$ and $y=0$, imply $c_2=\frac{l\beta} {1-\beta^2}$. Therefore the equation of the path of a missile directed at every instant towards rectilinearly moving target is given by 

\begin{equation}
  y = \frac{1}{2}[\frac{l^{-\beta}(l-x)^{1+\beta}}{1+\beta} - \frac{l^{\beta}(l-x)^{1-\beta}}{1-\beta}] + \frac{l\beta} {1-\beta^2}.
\label{seqn9}
\end{equation}
\noindent

    Let $T$ be the time taken by the missile to hit the target after leaving the station. Then when $x=l$, $t=T$ and Eq.(1) provides

\begin{equation}
  uT = y(x=l). 
\label{seqn10}
\end{equation}
\noindent
From Eq.(9), $y(x=l) = \frac{l\beta} {1-\beta^2}$ and therefore  

\begin{equation}
  T = \frac{y(x=l)}{u} = \frac{l\beta} {u(1-\beta^2)} = \frac{lv} {v^2-u^2}
\label{seqn11}
\end{equation}
\noindent
using $\beta=\frac{u}{v}$. The above equation tells that $v>u$ for the missile to hit the target.


\noindent
\begin{thebibliography}{99}

\bibitem{Sc55} L. I. Schiff, Quantum Mechanics, McGraw-Hill Kogakusha Ltd. (1955). 

\bibitem{Go50} H. Goldstein, Classical Mechanics, Addison-Wesley Publishing Company Inc. (1950). 

\end{thebibliography}
\end{document}